\PassOptionsToPackage{unicode}{hyperref}
\PassOptionsToPackage{hyphens}{url}
\PassOptionsToPackage{dvipsnames,svgnames,x11names}{xcolor}
\documentclass[
  12pt]{article}

\usepackage{amsmath,amssymb}
\usepackage{bm}
\usepackage{iftex}
\ifPDFTeX
  \usepackage[T1]{fontenc}
  \usepackage[utf8]{inputenc}
  \usepackage{textcomp} 
\else 
  \usepackage{unicode-math}
  \defaultfontfeatures{Scale=MatchLowercase}
  \defaultfontfeatures[\rmfamily]{Ligatures=TeX,Scale=1}
\fi
\usepackage{lmodern}
\ifPDFTeX\else  
\fi
\IfFileExists{upquote.sty}{\usepackage{upquote}}{}
\IfFileExists{microtype.sty}{
  \usepackage[]{microtype}
  \UseMicrotypeSet[protrusion]{basicmath} 
}{}
\makeatletter
\@ifundefined{KOMAClassName}{
  \IfFileExists{parskip.sty}{%
    \usepackage{parskip}
  }{
    \setlength{\parindent}{0pt}
    \setlength{\parskip}{6pt plus 2pt minus 1pt}}
}{
  \KOMAoptions{parskip=half}}
\makeatother
\usepackage{xcolor}
\makeatletter
\ifx\paragraph\undefined\else
  \let\oldparagraph\paragraph
  \renewcommand{\paragraph}{
    \@ifstar
      \xxxParagraphStar
      \xxxParagraphNoStar
  }
  \newcommand{\xxxParagraphStar}[1]{\oldparagraph*{#1}\mbox{}}
  \newcommand{\xxxParagraphNoStar}[1]{\oldparagraph{#1}\mbox{}}
\fi
\ifx\subparagraph\undefined\else
  \let\oldsubparagraph\subparagraph
  \renewcommand{\subparagraph}{
    \@ifstar
      \xxxSubParagraphStar
      \xxxSubParagraphNoStar
  }
  \newcommand{\xxxSubParagraphStar}[1]{\oldsubparagraph*{#1}\mbox{}}
  \newcommand{\xxxSubParagraphNoStar}[1]{\oldsubparagraph{#1}\mbox{}}
\fi
\makeatother

\usepackage{longtable,booktabs,array}
\usepackage{calc} 
\usepackage{etoolbox}
\makeatletter
\patchcmd\longtable{\par}{\if@noskipsec\mbox{}\fi\par}{}{}
\makeatother
\IfFileExists{footnotehyper.sty}{\usepackage{footnotehyper}}{\usepackage{footnote}}
\makesavenoteenv{longtable}
\usepackage{graphicx}
\makeatletter
\def\maxwidth{\ifdim\Gin@nat@width>\linewidth\linewidth\else\Gin@nat@width\fi}
\def\maxheight{\ifdim\Gin@nat@height>\textheight\textheight\else\Gin@nat@height\fi}
\makeatother
\setkeys{Gin}{width=\maxwidth,height=\maxheight,keepaspectratio}
\makeatletter
\def\fps@figure{htbp}
\makeatother

\makeatletter
\@ifpackageloaded{caption}{}{\usepackage{caption}}
\AtBeginDocument{%
\ifdefined\contentsname
  \renewcommand*\contentsname{Table of contents}
\else
  \newcommand\contentsname{Table of contents}
\fi
\ifdefined\listfigurename
  \renewcommand*\listfigurename{List of Figures}
\else
  \newcommand\listfigurename{List of Figures}
\fi
\ifdefined\listtablename
  \renewcommand*\listtablename{List of Tables}
\else
  \newcommand\listtablename{List of Tables}
\fi
\ifdefined\figurename
  \renewcommand*\figurename{Figure}
\else
  \newcommand\figurename{Figure}
\fi
\ifdefined\tablename
  \renewcommand*\tablename{Table}
\else
  \newcommand\tablename{Table}
\fi
}
\@ifpackageloaded{float}{}{\usepackage{float}}
\floatstyle{ruled}
\@ifundefined{c@chapter}{\newfloat{codelisting}{h}{lop}}{\newfloat{codelisting}{h}{lop}[chapter]}
\floatname{codelisting}{Listing}

\makeatother
\makeatletter
\@ifpackageloaded{caption}{}{\usepackage{caption}}
\@ifpackageloaded{subcaption}{}{\usepackage{subcaption}}
\makeatother

\ifLuaTeX
  \usepackage{selnolig}  
\fi
\usepackage[]{natbib}
\usepackage{bookmark}

\IfFileExists{xurl.sty}{\usepackage{xurl}}{} 
\hypersetup{
  pdftitle={Title},
  pdfauthor={Author 1; Author 2},
  pdfkeywords={3 to 6 keywords, that do not appear in the title},
  colorlinks=true,
  linkcolor={blue},
  filecolor={Maroon},
  citecolor={Blue},
  urlcolor={Blue},
  pdfcreator={LaTeX via pandoc}}

\newcommand{\anon}{1}

\begin{document}

\def\spacingset#1{\renewcommand{\baselinestretch}%
{#1}\small\normalsize} \spacingset{1}

\date{}

\if1\anon
{
  \title{\bf Spatial Capture-Recapture With Penalized Regression Splines to Flexibly Model Wildlife Density and Distribution}
  \author{Andrew E. Seaton\thanks{
    The authors gratefully acknowledge funding from the Royal Society of New Zealand through Marsden Grants UOA-1929 and UOA-2321.}\hspace{.2cm}\\
    School of Mathematics and Statistics, University of Glasgow \\ United Kingdom \\
    \\
    David L. Borchers \\
    School of Mathematics and Statistics, University of St Andrews \\ United Kingdom \\
    \\
    Milou Groenenberg\thanks{Present address: Copernicus Institute of Sustainable Development, Utrecht University, the Netherlands} \\
    World Wide Fund for Nature, Phnom Penh \\ Cambodia \\
    \\
    Ben C. Stevenson \\
    Department of Statistics, University of Auckland \\ New Zealand
  }
  \maketitle
} \fi

\if0\anon
{
  \bigskip
  \bigskip
  \bigskip
  \begin{center}
    {\LARGE\bf Spatial Capture-Recapture With Penalized Regression Splines to Flexibly Model Wildlife Density and Distribution}
\end{center}
  \medskip
} \fi

\bigskip
\begin{abstract}
Spatial capture-recapture models are routinely used to estimate the abundance and distribution of wild animal populations and involve a latent spatial point process of animal activity centres that describes the spatial distribution of individuals. While traditional spatial capture-recapture models use a Poisson process, the assumption of conditional independence between points is often violated in practice due to factors not included in the point process, such as social clustering, territoriality, or preferential selection of habitat due to unobserved covariates. Log-Gaussian Cox processes are commonly used in spatial statistics to overcome weaknesses of Poisson processes, but methods to fit them within spatial capture-recapture do not currently exist. Here, we present a spatial capture-recapture framework that allows for the use of penalized regression splines to describe the activity centre distribution, with model fitting via a Laplace-approximate penalized marginal maximum likelihood approach. Our method approximates using a log-Gaussian Cox process for activity centres, and allows flexible modelling of nonlinear effect of covariates on density. We illustrate the use of our method with a simulation study and two case-studies. We demonstrate that, while population size estimates of traditional approaches are robust to density model misspecification, our approach substantially improves the estimation of spatial animal distributions.
\end{abstract}

\noindent%
\vfill

\newpage
\spacingset{1.8} 

\section{Introduction}

Spatial capture-recapture (SCR) methods estimate the abundance and spatial distribution of animals \citep{efford_firstsecr_2004, borchers_SpatiallyExplicitMaximum_2008, royle_HierarchicalModelSpatial_2008}, and are now a popular statistical tool used to understand wildlife populations \citep{tourani_SCRReview_2022}. SCR models probabilistically describe two key processes: a latent spatial point process for animals' activity centres (the centroids of individual animal movement during the survey), and a process for detections of animals at detectors (such as camera traps), conditional on their locations. The activity centres are not observed, however, which presents challenges for model fitting: existing approaches involve integrating over all possible activity centre locations for every detected animal \citep{borchers_SpatiallyExplicitMaximum_2008} or sampling the locations within a Markov chain Monte Carlo algorithm \citep{royle_HierarchicalModelSpatial_2008}.

So far, almost every proposed SCR model has used either a binomial or Poisson process to model activity centres, and both involve assumptions of independence between their locations---a step that considerably simplifies model fitting because we can can separately integrate over, or sample, a location for each individual animal without considering the others. Often applications use homogeneous versions of these processes \citep{borchers_SCRReview_2015}, the simplest choice, although in many contexts ecologists require inference into spatiotemporal changes in animal density. For example, key inferential targets may include species-habitat associations \citep{rather_DensityEstimationTiger_2021, gaukler_InvestigatingEffectsSoil_2020, mann_LeopardFavouriteSpots_2020, furnasOverabundanceBlackTailedDeer2020, welfeltFactorsAssociatedBlack2019}, spatially-varying population dynamics \citep{broekhuis_ResourcePulsesInfluence_2021}, the effect of management policies \citep{toblerResponsiblyManagedLogging2018}, responses to disturbance and stressors \citep{clare_SatellitedetectedForestDisturbance_2019, sutherland_LargescaleVariationDensity_2018, laufenbergCompoundingEffectsHuman2018}, and identification of areas of key conservation importance \citep{guptaReserveDesignOptimize2019, penjorIdentifyingImportantConservation2018}.

\cite{borchers_SpatiallyExplicitMaximum_2008} and \cite{zhang_ihp_2023} considered SCR models with inhomogeneous Poisson processes, which allow inference into spatiotemporal changes in animal density. For these models, intensity of activity centres varies according to the log-linear effect of spatial covariates: $\log\{D(\bm{s})\} = \beta_0 + \sum_j \beta_j x_j(\bm{s})$. Here, $D(\bm{s})$ is the intensity of activity centres (i.e., animal population density) and $x_j(\bm{s})$ is the $j$th covariate at location $\bm{s}$. However, one key assumption of the inhomogeneous Poisson process is that the locations of animals are assumed to be conditionally independent given the intensity. We therefore rely on observing covariates that explain the spatial variation in population density and modelling them appropriately. Poor inference about animal distribution can result if the effects of covariates on $\log\{D(\bm{s})\}$ are not linear, if there is clustering due to covariates that have not been observed, or if there is clustering due to mechanisms that are not described by environmental covariates (e.g., flocking or herding). Incorrectly specified models suffer from bias, underestimation of uncertainty, and spurious significance for fixed effects estimates.

In spatial statistics, a popular extension to Poisson processes is the log-Gaussian Cox process (LGCP) and accounts for clustering of locations that cannot be explained by available covariates, but so far has not been used within an SCR model. In lieu of using an LGCP, the prevailing approach to flexibly model spatiotemporal changes in density within SCR is to use unpenalized regression splines \citep{borchers_scr_splines_2014}, allowing nonlinear effects of covariates on $\log\{D(\bm{s})\}$, or spatial smoothing over x- and y-coordinates. Unpenalized regression splines require the analyst to specify smoothness, for example by selecting the number of basis functions alongside the location of knots. One approach to choosing a level of smoothness is to fit models with different smoothness settins and select one using information criteria such as AIC; an example is present in the documentation for the \texttt{secr} package (see the vignette on modelling density surfaces).

The use of regression splines in statistical modelling has been popularized through generalized additive models \citep[GAMs;][]{wood_GeneralizedAdditiveModels_2017} and widely used R packages such as \texttt{mgcv}. Statistical methodology and software for GAMs have reached a point where, in applied work, unpenalized splines are largely eschewed in favour of penalized splines, for which the number of basis functions is fixed and the degree of smoothness can be estimated by including a `wiggliness' penalty in the likelihood. Advantages of penalized splines include the following: (1) inference from unpenalized splines with wiggliness selected by AIC does not accommodate uncertainty in the specified smoothness, but it is possible to do so when smoothness is estimated within a single model \citep[][p.~302]{wood_GeneralizedAdditiveModels_2017}; (2) penalized regression splines are robust to knot placement, whereas unpenalized are not \citep[][pp.~166--167]{wood_GeneralizedAdditiveModels_2017}; and (3) using AIC to select smoothness risks severe undersmoothing \citep{wood_FastStableRestricted_2011}, which can be avoided using penalized splines. Indeed, in our own experiences with unpenalized splines in SCR models, some data sets produce AIC values that continue to decrease well past any reasonable level of smoothness, resulting in clear overfitting. However, to date, no methods are available to fit penalized splines within SCR.

In this paper, we present the first implementation of penalized regression splines for the point process model in SCR. A further advantage of penalized regression splines is that they allow use of LGCPs rather than Poisson processes to model activity centres: an LGCP extends an inhomogeneous Poisson process by adding a latent Gaussian random field, and can be approximated using basis functions and fitted using penalized regression splines \citep{dovers_lgcp_2024, yueBayesianAdaptiveSmoothing2014} and other structured random effects \citep{yuan_PointProcessModels_2017, simpson_GoingGridComputationally_2016}. Although \cite{reich_SpatialCapturerecaptureModel_2014} and \cite{diana_VectorPointProcesses_2022a} have developed SCR models capable of modelling negative spatial correlation between activity centre locations, all other SCR models that have been proposed involve an assumption of independence between locations via binomial or Poisson point processes. None exist to model positive spatial correlation caused by effects that cannot be explained by available covariates, but this can be achieved with an LGCP.

The use of penalized regression splines within SCR therefore offers a promising route towards using more flexible spatial processes to model the dependence between locations of activity centres. One unique challenge relative to other applications of LGCPs is that here locations are latent: our goal is to fit a complex spatial point process without observing the point pattern. Our approach involves a Laplace-approximate penalized marginal maximum likelihood implementation, leveraging automatic differentiation via the TMB package \citep{kristensen_TMBAutomaticDifferentiation_2016}. Our method allows the fitting of nonlinear effects of spatial covariates on animal density, and for the use of LGCPs within SCR to describe spatial clustering that cannot be explained by these covariates. We present two applications of our method to wildlife data, alongside a simulation study.

\section{Spatial capture-recapture}
\label{sec:scr}

Under an SCR model, we assume animal activity centres are generated by a point process within the study area $A$, where $\bm{s}_i \in A$ containins coordinates of the $i$th individual's activity centre. Depending on the point process, the total number of animals, $N$, could either be fixed (e.g., for a binomial point process) or random (for a Poisson process). The detection data for the $i$th animal are contained in $\bm{\omega}_i$.

For simplicity, in this section we focus on a single-occasion survey involving binary proximity detectors \citep{efford_ProximityDetectors_2009}, although the key ideas in this paper are applicable to all varieties of SCR surveys; see \cite{borchers_SCRReview_2015} for an overview. For binary proximity-detector surveys, researchers deploy $J$ detectors at fixed locations. The detection data for the $i$th animal is a vector of binary responses, $\bm{\omega}_i = (\omega_{i1}, \cdots, \omega_{iJ})$, where $\omega_{ij} = 1$ if the $i$th individual was detected by the $j$th detector, and $\omega_{ij} = 0$ otherwise. The probability that an animal is detected by a detector is governed by a detection function, and depends on the distance between the activity centre and detector locations. A commonly used detection function is the half-normal, given by $g(d) = g_0\exp(-d^2/2\sigma^2)$, where $g_0 \in (0,1)$ and $\sigma > 0$ are model parameters, and $d$ is the distance between the activity centre and the detector.

In this paper, we use the function $f$ for probability mass functions (PMFs) and probability density functions (PDFs), using subscripts to denote its variable. We condition on random effects but, to avoid cumbersome notation, suppress parameters. Of the $N$ total animals, only $n$ are observed: those detected by at least one detector. The probability that an animal with activity centre $\bm{s}$ is detected by at least one detector is given by $p_\cdot(\bm{s}) = 1 - \prod_{j = 1}^J [1 - g\{d_j(\bm{s})\}]$. An observed $\bm{\omega}_i$ is zero-truncated because it is not possible to observe $\bm{\omega}_i = (0, \cdots, 0)$. Its PMF, conditional on the activity centre, is therefore $f_{\bm{\omega} \mid \bm{s}}(\bm{\omega}_i \mid \bm{s}_i) = \prod_{j = 1}^J g\{d_j(\bm{s}_i)\}^{\omega_{ij}} [1 - g\{d_j(\bm{s}_i)\}]^{(1 - \omega_{ij})}/p_\cdot(\bm{s}_i)$, where each term in the product is a Bernoulli PMF, one for each detector, and the denominator accounts for zero-truncation.

Because activity centres are not observed, the likelihood function for SCR models requires integrating over the activity centre locations. In general, regardless of survey type, the likelihood has the form
\begin{equation}
  \mathcal{L}(\bm{\theta}) = \int_{A^n} f_{\bm{\Omega} \mid \bm{S}}(\bm{\Omega} \mid \bm{S}) f_{\bm{S}}(\bm{S}) \, \mathrm{d}\bm{S}, \label{eq:general-scr-likelihood}
\end{equation}
where $\bm{\Omega} = (\bm{\omega}_1, \cdots, \bm{\omega}_n)$, $\bm{S} = (\bm{s}_1, \cdots, \bm{s}_n)$ contain all detected animals' detection data and activity centre coordinates, and $A^n$ is the set of all possible combinations of locations for the $n$ animals, a $2n$-dimensional space. In general, this $2n$-dimensional integral is substantially challenging to evaluate or approximate.

However, existing SCR approaches typically use binomial or Poisson point processes for $\bm{S}$, which imply independence between locations and allow the unwieldy $2n$-dimensional integral to be separated into $n$ two-dimensional integrals. For example, if activity centres arise from a Poisson process with intensity function $D(\bm{s})$, then activity centres for detected animals arise from a Poisson process with intensity function $D(\bm{s})p_\cdot(\bm{s})$ \citep{borchers_SCRReview_2015}. Using the likelihood function for an inhomogeneous Poisson process \citep[][p.\ 121]{illian_spatialPointProcesses}, we obtain $f_{\bm{S}}(\bm{S}) = \exp(-\Lambda) \prod_{i = 1}^n \{ D(\bm{s}_i) p_\cdot(\bm{s}_i) \}$. Here, $\Lambda = \int_A D(\bm{s})p_\cdot(\bm{s}) \mathrm{d}\bm{s}$ is the expected number of detected animals, where the integral is approximated numerically. Subsequently, we can rewrite the likelihood as
\begin{align}
  \mathcal{L}(\bm{\theta}) &= \int_{A^n} f_{\bm{\Omega} \mid \bm{S}}(\bm{\Omega} \mid \bm{S}) \exp(-\Lambda) \prod_{i = 1}^n \left \{ D(\bm{s}_i) p_\cdot(\bm{s}_i) \right\}  \, \mathrm{d}\bm{S} \nonumber \\
  &= \exp(-\Lambda) \int_{A^n} \prod_{i = 1}^n f_{\bm{\omega} \mid \bm{s}}(\bm{\omega}_i \mid \bm{s}_i) \prod_{i = 1}^n \left \{ D(\bm{s}_i) p_\cdot(\bm{s}_i) \right\}  \, \mathrm{d}\bm{S} \nonumber \\
  &=  \exp(-\Lambda) \prod_{i = 1}^n \int_A f_{\bm{\omega} \mid \bm{s}}(\bm{\omega}_i \mid \bm{s}_i) D(\bm{s}_i)p_\cdot(\bm{s}_i) \, \mathrm{d}\bm{s}_i, \label{eq:scr-separable-integral}
\end{align}
with separability possible because the integrand factors into a product with terms that each depend on a different activity centre. This likelihood function is far more manageable because we have $n$ two-dimensional integrals rather than a $2n$-dimensional integral.

Shifting away from binomial or Poisson point processes for $\bm{S}$ is challenging. For example, an LGCP is an extension of an inhomogeneous Poisson process for which $\log \{D(\bm{s})\}$ is a Gaussian process, but the marginal PDF $f_{\bm{S}}(\bm{S})$ is not available in closed form. Even for some specified collection of activity centres $\bm{S}$, we may not easily be able to compute or approximate $f_{\bm{S}}(\bm{S})$, and, to further complicate matters, the likelihood in Eq \eqref{eq:general-scr-likelihood} requires doing so for all possible configurations of activity centres. Moreover, the activity centres' dependence on a latent Gaussian random field makes it impossible to respecify the $2n$-dimensional integral as $n$ separate two-dimensional integrals as we did, above, for the SCR model with a Poisson process for activity centres, further complicating the computation or approximation of the likelihood function.

Here, by leveraging links between Gaussian processes and smoothing \citep{dovers_lgcp_2024, miller_spdeSmoothing_2020, yueBayesianAdaptiveSmoothing2014}, we develop a method using penalized regression splines to fit SCR models with an LGCP for activity centres. The method can additionally be used to model nonlinear effects of environmental covariates on $D(\bm{s})$. We introduce necessary background details about penalized regression splines below, and how they relate to LGCPs.

\section{Penalized regression splines and LGCPs}
\label{sec:prs}

\subsection{An overview of penalized regression splines}
\label{sec:prs-overview}

Regression splines allow analysts to fit smooth effects of covariates. In this section, we consider the simple case of a model with a linear effect of a single variable, $x$, and a nonlinear effect modelled using a spline for another variable, $z$:
\begin{equation}
  g(\mu_i) = \beta_0 + \beta_1 x_i + \sum_{b = 1}^B u_b \phi_b(z_i). \label{eq:general-spline}
\end{equation}
Here, $\mu_i$ is a parameter of the distribution for the $i$th response, $y_i$, and $g$ is a link function. The regression spline $f(z) = \sum_{b = 1}^B u_b \phi_b(z)$ describes a smooth effect of $z$ and is represented by $B$ basis functions, the $b$th of which is $\phi_b(z)$. The basis functions have coefficients $\bm{u} = (u_1, \cdots, u_B)$. The distribution of $y_i$ may depend on other parameters, $\bm{\theta}$. For example if $y_i \sim \mathrm{Normal}(\mu_i, \sigma^2)$, we must also estimate $\sigma$.

The basis functions are generated depending on the choice of spline. Examples include B-splines, where basis functions have local support (i.e., they are nonzero only between some finite number of knots, the locations of which are selected by the analyst) and thin plate regression splines, for which the generation of the basis functions is rather more complicated \cite[see][pp.~216--219, for further details]{wood_GeneralizedAdditiveModels_2017}. One way to fit an unpenalized spline is to maximize the log-likelihood, $\log\{\mathcal{L}(\bm{\beta}, \bm{u}, \bm{\theta})\} = \log\{f_{\bm{y}}(\bm{y})\}$, over $\bm{\beta}$, $\bm{u}$, and $\bm{\theta}$, where $f_{\bm{y}}(\bm{y})$ is the joint PDF of the response. In this case, the smoothness of the spline, $f(z)$, is determined by $B$ and the basis functions, selected by the analyst.

For penalized splines we introduce a penalty, so that we maximize $\log\{\mathcal{L}(\bm{\beta}, \bm{u}, \bm{\theta})\} - \lambda W(\bm{u})$ instead of $\log\{\mathcal{L}(\bm{\beta}, \bm{u}, \bm{\theta})\}$. As we will show below, the penalty approach is equivalent to considering $\bm{u}$ as random, so in our notation we begin conditioning on $\bm{u}$ in functions where $\bm{u}$ is considered fixed. The penalty $W(\bm{u})$ measures the wiggliness of $f(z \mid \bm{u})$. A common measure is the integrated squared derivative $W(\bm{u}) = \int f^{\prime \prime}(z \mid \bm{u}) \mathrm{d}z$. Intuitively, the smoothest possible function is linear, for which $f^{\prime \prime}(z \mid \bm{u}) = 0$ and therefore has zero wiggliness under this measure. If $f(z \mid \bm{u})$ fluctuates substantially then $f^{\prime \prime}(z \mid \bm{u})$ will be large at some values of $z$, thereby giving a large value for $W(\bm{u})$. The $\lambda$ parameter therefore controls smoothness: if $\lambda = 0$ then there is no penalty, and the spline has the flexibility to be as wiggly as it needs to be in order to maximize the unpenalized likelihood, but if $\lambda$ is large then the spline must be smooth to minimize the penalty $\lambda W(\bm{u})$ and therefore maximize $\log\{\mathcal{L}_p(\bm{\beta}, \bm{u}, \bm{\theta})\}$.

An alternative but equivalent approach uses ideas from mixed-effects models, where we treat $\bm{u}$ as a vector of random effects \citep{wood_FastStableRestricted_2011}. The motivation for this is the fact that, under a specific choice of basis functions, $W(\bm{u}) = \bm{u} \bm{Q} \bm{u}$, where $\bm{Q}_{ij}$ is the smoothing penalty applied to $\phi_i(z) \phi_j(z)$.  This is proportional to a Gaussian density with precision $\lambda$ and precision matrix $\bm{Q}$. Under a reparameterization of basis functions, we can assume the random effects are independent and identically normally distributed with common variance, $u_b \sim \mathrm{Normal}(0, \tau^2)$, where $\tau$ is the standard deviation of $u_b$. To evaluate the likelihood function, we must integrate over the random effects $\bm{u}$:
\begin{equation}
  \mathcal{L}(\bm{\beta}, \bm{\theta}, \tau) = \int_{{\mathbb{R}^B}} f_{\bm{y} \mid \bm{u}}(\bm{y} \mid \bm{u}) f_{\bm{u}}(\bm{u}) \, \mathrm{d}\bm{u}. \label{eq:prs-likelihood}
\end{equation}
\cite{wood_FastStableRestricted_2011} proposed using the Laplace approximation to evaluate this integral. Model fitting can be achieved by maximizing this marginalized likelihood over $\bm{\beta}$, $\bm{\theta}$, and $\tau$. Under this mixed-effects formulation, $\tau$ replaces $\lambda$ to control the smoothness of the spline: if $\tau = 0$ then all coefficients $\bm{u}$ will be zero, thus $f(z) = 0$ and we have no wiggliness (i.e., maximum smoothness). The larger $\tau$, the more wiggliness we allow, giving a smoother spline.

\subsection{A link between penalized splines and LGCPs}
\label{sec:spline-lgcp-link}

The intensity function of an inhomogeneous Poisson process is controlled by fixed effects of spatial covariates, $\log\{D(\bm{s})\} = \beta_0 + \sum_j \beta_j x_j(\bm{s})$. To introduce LGCPs, we consider an extension of an inhomogeneous Poisson process, where we include a zero-mean Gaussian random field, $\xi(\bm{s})$, to model additional spatial variation in the intensity function that cannot be explained by the fixed effects. So, an LGCP over survey region $A$ with a single fixed effect for the spatial covariate $x(\bm{s})$ has intensity function
\begin{equation}
  \log\{D(\bm{s} \mid \xi)\} = \beta_0 + \beta_1 x(\bm{s}) + \xi(\bm{s}). \label{eq:lgcp-intensity}
\end{equation}
Because $\xi(\bm{s})$ is a random variable for any location $\bm{s} \in A$, a Gaussian random field is a collection of infinite random variables. A zero-mean Gaussian random field is fully characterized by its covariance function, and a common approach is to model the covariance using the distance between locations: $\mathrm{Cov}\{\xi(\bm{s}), \xi(\bm{s}^*)\} = C\{d(\bm{s}, \bm{s}^*)\}$, where the analyst selects some functional form for $C(d)$ involving parameters $\bm{\gamma}$, a vector containing the covariance function parameters. In our notation $D(\bm{s} \mid \xi)$, we explicitly acknowledge that the intensity function depends on the random field by placing $\xi$ after a conditioning bar.

Conditional on the Gaussian random field, $\bm{S} = (\bm{s}_1, \cdots, \bm{s}_n)$ are a realisation from an inhomogeneous Poisson process, and thus have the conditional PDF $f_{\bm{S}}(\bm{S} \mid \xi) = \exp\{-\Lambda(\xi)\} \prod_{i = 1}^n D(\bm{s}_i \mid \xi)$, where $\Lambda(\xi) = \int_A D(\bm{s} \mid \xi) \mathrm{d}\bm{s}$ can be approximated numerically. However, becuase we do not observe $\xi$, the marginal likelihood involves an intractable infinite-dimensional integral,
\begin{equation}
  \mathcal{L}(\bm{\bm{\beta}}, \bm{\gamma}) = \int f_{\bm{S}}(\bm{S} \mid \xi) f_\xi(\xi) \mathrm{d}\xi = \int \exp\{-\Lambda(\xi)\} \prod_{i = 1}^n D(\bm{s} \mid \xi) f_\xi(\xi) \mathrm{d}\xi. \label{eq:lgcp-intractable}
\end{equation}

One strategy is to approximate the Gaussian random field using two-dimensional basis functions, $\xi(\bm{s}) \approx \sum_{b = 1}^B u_b \phi_b(\bm{s})$. To ensure this approximation adheres to the random nature of $\xi(\bm{s})$, we treat the coefficients $\bm{u}$ as random, where, under a sensible selection of basis functions, we may assume the coefficients are independent with distribution $u_b \sim \mathrm{Normal}(0, \tau^2)$ \citep{dovers_falgcp_2023}. Our approximating model becomes
\begin{equation}
  \log\{D(\bm{s} \mid \bm{u}) \} = \beta_0 + \beta_1 x(\bm{s}) + \sum_{b = 1}^B u_b \phi_b(\bm{s}), \label{eq:lgcp-approx-intensity}
\end{equation}
which bears a striking resemblence to Equation \eqref{eq:general-spline}, including the linear combination of basis functions with random, normally distributed coefficients $\bm{u}$. Indeed, the only fundamental difference between this model and that described in Section \ref{sec:prs-overview} is that here the equation describes the intensity function for a point process in continuous two-dimensional space, rather than $\mu_i$, a parameter that controls the distribution of the response $y_i$.

Fitting (approximations to) LGCPs can be achieved using similar strategies to penalized regression splines: we use a mixed-effects model approach, providing the likelihood
\begin{equation}
  \mathcal{L}(\bm{\beta}, \bm{\theta}, \tau) = \int_{{\mathbb{R}^B}} f_{\bm{S}}(\bm{S} \mid \bm{u}) f_{\bm{u}}(\bm{u}) \mathrm{d}\bm{u} = \int_{{\mathbb{R}^B}} \exp\{-\Lambda(\bm{u})\} \prod_{i = 1}^n D(\bm{s} \mid \bm{u}) f_{\bm{u}}(\bm{u}) \, \mathrm{d}\bm{u}. \label{eq:lgcp-likelihood}
\end{equation}
The randomness of the approximate Gaussian random field is governed by the $B$ random effects $\bm{u}$, replacing $\xi$, so we now condition on and integrate over $\bm{u}$ rather than $\xi$. Computing the likelihood is now manageable because we have reduced the dimensionality of the integral from infinite to $B$. Similar to the mixed-effects approach to smoothing splines, the Laplace approximation can be used for the integral over $\bm{u}$ \citep{dovers_falgcp_2023, yuan_PointProcessModels_2017, simpson_GoingGridComputationally_2016}. Due to the similarities between Equations \eqref{eq:general-spline} and \eqref{eq:lgcp-intensity}, and between \eqref{eq:prs-likelihood} and \eqref{eq:lgcp-likelihood}, it is perhaps unsurprising that existing software for penalized regression splines can also be used to fit LGCPs \citep{dovers_lgcp_2024}.  However, in the SCR context, the point pattern is latent and so we cannot use off-the-shelf GLM-type methods or software.

\section{Introducing penalized regression splines to SCR}
\label{sec:scr-with-splines}

We propose introducing penalized regression splines to SCR via the intensity function:
\begin{equation}
  \log\{D(\bm{s} \mid \bm{u})\} = \beta_0 + \sum_{j = 1}^J \beta_j x_j(\bm{s}) + \sum_{k = 1}^K \sum_{b = 1}^{B_k}  u_{kb} \phi_{kb}\{z_k(\bm{s})\}. \label{eq:scr-lgcp-intensity}
\end{equation}
Here, we have made an extension to Equations \eqref{eq:general-spline} and \eqref{eq:lgcp-intensity} by allowing $J$ total fixed-effects terms, and separate splines for $K$ total other variables. The $k$th spline models a smooth effect of the spatial covariate $z_k(\bm{s})$ and has $B_k$ basis functions, the $b$th of which is $\phi_{kb}(z_k)$ and has coefficient $u_{kb}$. The total number of basis functions is $B = \sum_{k = 1}^K B_k$. The vector $\bm{u} = (u_{11}, \cdots, u_{1B_1}, u_{21}, \cdots, u_{KB_K})$ has $B$ elements, containing the random coefficients for all $K$ splines. Each spline is a smooth function over a different variable, each requiring its own level of smoothing, so we use a different $\tau$ parameter for each: $u_{kb} \sim \mathrm{Normal}(0, \tau_k^2)$. Like we did in Section \ref{sec:spline-lgcp-link}, we acknowledge that the intensity function depends on the random vector $\bm{u}$ via the notation $D(\bm{s} \mid \bm{u})$. The fixed effects are consistent with existing SCR models involving inhomogeneous Poisson processes \citep{borchers_SpatiallyExplicitMaximum_2008, zhang_ihp_2023}, while the penalized regression splines are the key novelty of this paper.

If the $k$th spline describes smooth effects over space, then $z_k(\bm{s}) = \bm{s}$. In this case, the intensity function appears in the same form as Equation \eqref{eq:lgcp-approx-intensity}, and therefore activity centres arise from an LGCP. Therefore, our approach allows us to accommodate dependence between locations and model spatial variation in density above and beyond what is explained by the fixed effects. Instead of defining a Gaussian random field over $\bm{s}$, or in addition, we can define them over covariates to model smooth, nonlinear effects on log-density, in which case $z_k(\bm{s})$ is the value of the covariate at location $\bm{s}$.

Under this model, the marginal distribution of detected animals' activity centres, $f_{\bm{S}}(\bm{S})$ is not that from an inhomogeneous Poisson process, because its intensity depends on the random vector $\bm{u}$. Conditional on $\bm{u}$, however, we revert to the standard SCR scenario where $\bm{S}$ is a point pattern generated by an inhomogeneous Poisson process with intensity $D(\bm{s} \mid \bm{u}) p_\cdot(\bm{s})$ \citep{borchers_SCRReview_2015}. Therefore $f_{\bm{S} \mid \bm{u}}(\bm{S} \mid \bm{u})$ is the usual inhomogeneous Poisson process likelihood \citep[][p.\ 121]{illian_spatialPointProcesses}, giving
\begin{equation*}
  f_{\bm{S} \mid \bm{u}}(\bm{S} \mid \bm{u}) =  \exp\{-\Lambda(\bm{u})\} \prod_{i = 1}^n D(\bm{s}_i \mid \bm{u}) p_\cdot(\bm{s}_i).
\end{equation*}
The key changes from existing SCR models (Section \ref{sec:scr}) is that the expected number of detected animals, $\Lambda(\bm{u}) = \int_A D(\bm{s} \mid \bm{u})p_\cdot(\bm{s}) \, \mathrm{d}\bm{s}$, and the intensity function for all animals, $D(\bm{s} \mid \bm{u})$, both depend on random effects $\bm{u}$. The marginal PDF for detected animals' activity centres is $f_{\bm{S}}(\bm{S}) = \int_{\mathbb{R}^B} f_{\bm{S} \mid \bm{u}}(\bm{S} \mid \bm{u}) f_{\bm{u}}(\bm{u}) \, \mathrm{d}\bm{u}$, where $f_{\bm{u}}(\bm{u}) = \prod_{k = 1}^K \prod_{b = 1}^{B_k} f_u(u_{kb})$, due to independence between basis function coefficients that arises from a sensible construction of basis functions \citep{dovers_falgcp_2023}. Here, $f_u(u_{kb})$ is the PDF of $u_{kb} \sim \mathrm{Normal}(0, \tau_k^2)$.

We develop our model's likelihood function starting from Equation \eqref{eq:general-scr-likelihood}, which provides
\begin{align}
  \mathcal{L}(\bm{\theta}) &= \int_{A^n} f_{\bm{\Omega} \mid \bm{S}}(\bm{\Omega} \mid \bm{S}) f_{\bm{S}}(\bm{S}) \, \mathrm{d}\bm{S} \nonumber \\
                           &= \int_{A^n} f_{\bm{\Omega} \mid \bm{S}}(\bm{\Omega} \mid \bm{S}) \int_\mathbb{R^B} f_{\bm{S} \mid \bm{u}}(\bm{S} \mid \bm{u}) f_{\bm{u}}(\bm{u}) \, \mathrm{d}\bm{u} \, \mathrm{d}\bm{S}, \nonumber \intertext{and, after switching the order of integration we obtain}
                           &= \int_{\mathbb{R}^B} f_{\bm{u}}(\bm{u}) \int_{A^n} f_{\bm{\Omega} \mid \bm{S}}(\bm{\Omega} \mid \bm{S}) f_{\bm{S} \mid \bm{u}}(\bm{S} \mid \bm{u}) \, \mathrm{d}\bm{S} \, \mathrm{d}\bm{u}. \nonumber \intertext{Upon conditioning on $\bm{u}$, the locations $\bm{S}$ arise from a Poisson process, thus the integral over activity centres is separable as per Equation \eqref{eq:scr-separable-integral}, giving}
                           &= \int_{\mathbb{R}^B} f_{\bm{u}}(\bm{u}) \exp\{-\Lambda(\bm{u})\} \prod_{i = 1}^n \int_{A} f_{\bm{\omega} \mid \bm{s}}(\bm{\omega}_i \mid \bm{s}_i) D(\bm{s}_i \mid \bm{u})p_\cdot(\bm{s}_i) \, \mathrm{d}\bm{s}_i \, \mathrm{d}\bm{u}. \label{eq:final-likelihood}
\end{align}

The inner integrals over activity centres can be handled in the standard way, by discretizing $A$ into cells and taking a Reimann sum \citep{borchers_SpatiallyExplicitMaximum_2008}. However, our method introduces the complication of an outer $B$-dimensional integral over $\bm{u}$. This integral is the cost of modelling spatial dependence between activity centres that cannot be explained by available environmental covariates: the likelihood is no longer a product of PDFs for independent random variables like it is for existing SCR models, Equation \eqref{eq:scr-separable-integral}.

To address this challenge, we propose the use of the Laplace approximation for the integration with respect to $\bm{u}$, following the approaches of \cite{wood_FastStableRestricted_2011} for penalized splines and \cite{dovers_falgcp_2023} for LGCPs. This approximation replaces the $B$-dimensional integration problem with a $B$-dimensional optimization problem, and operates as follows. First, the mode of the log-integrand is found by maximizing over $\bm{u}$, and the Hessian matrix of second derivatives with respect to $\bm{u}$ is computed. The approximation uses the integral of a Gaussian function that matches the mode and curvature, which is available in closed form.

We use \texttt{TMB} to compute the likelihood, an R package that is computationally efficient at fitting models involving Laplace-approximated likelihood functions. Based on a \texttt{C++} likelihood provided by the user, \texttt{TMB} implements automatic differentiation, which considerably speeds up optimization over $\bm{u}$ and allows fast computation of the Hessian while computing the Laplace approximation. Similarly, \texttt{TMB} speeds up optimization of $\mathcal{L}(\bm{\theta})$ over $\bm{\theta}$ to compute point estimates $\widehat{\bm{\theta}}$, and allows for fast computation of the Hessian of $\mathcal{L}(\bm{\theta})$ at $\widehat{\bm{\theta}}$ with respect to $\bm{\theta}$ to calculate standard errors. See \cite{kristensen_TMBAutomaticDifferentiation_2016} for further details about \texttt{TMB} and the role of the Laplace approximation and automatic differentiation for model-fitting.

\section{Software implementation}

We provide R functions to fit the model we propose as supplementary material. The primary function is \texttt{fit.scr.smooth()}. Arguments include (1) \texttt{capt}, the detection data $\bm{\Omega}$; (2) \texttt{traps}, the detector locations; (3) \texttt{mask}, mask point locations, a fine grid of points used to approximate spatial integrals for $\Lambda(\bm{u})$ and over activity centres in Equation \eqref{eq:final-likelihood}, as per \cite{borchers_SpatiallyExplicitMaximum_2008}; (3) \texttt{model}, a model formula specifying fixed effects, and effects to be fitted with penalized regression splines; and (5) \texttt{mask.df}, a data frame with spatial covariates measured at the mask point locations.

The \texttt{model} formula uses the same syntax as the popular \texttt{mgcv} package, and any type of spline implemented in \texttt{mgcv} is available (thin plate regression splines, P-splines, cyclic splines, and so on). The variables \texttt{x} and \texttt{y} are reserved for x- and y-coordinates. For example, to fit a fixed linear effect of variable \texttt{a}, a smooth effect of \text{b}, and a spatial smoother that gives rise to an LGCP, we could use \texttt{model = \string~ a + s(b, k = 10) + s(x, y, k = 32)}. Thin plate regression splines are used by default, where \texttt{k} provides the basis dimension, specifying maximum allowable wiggliness. Internally within our function, the \texttt{mgcv} function \texttt{smooth2random()} reparameterizes the basis functions so that the random coefficients for the $k$th spline are independently and identically distributed, $u_{kb} \sim \textrm{Normal}(0, \tau_k^2)$. Model fitting is achieved via a call to a compiled \texttt{TMB} executable. We include helper functions, which, for example, plot estimated density surfaces and estimated smooth functions.

\section{Applications}

Here we present two applications to illustrate the use of our model. The first involves data collected on a hair-snare survey of black bears, where we fit an LGCP via spatial smoothing over x- and y-coordinates. The second involves data collected on acoustic surveys of the southern yellow-cheeked crested gibbon, where we fit fixed effects of one spatial variable, and a smooth effect of another, in addition to spatial smoothing over x- and y-coordinates.

\subsection{Application to black bear hair-snare data}
\label{sec:bears}

\begin{figure}
	\includegraphics[width = \textwidth]{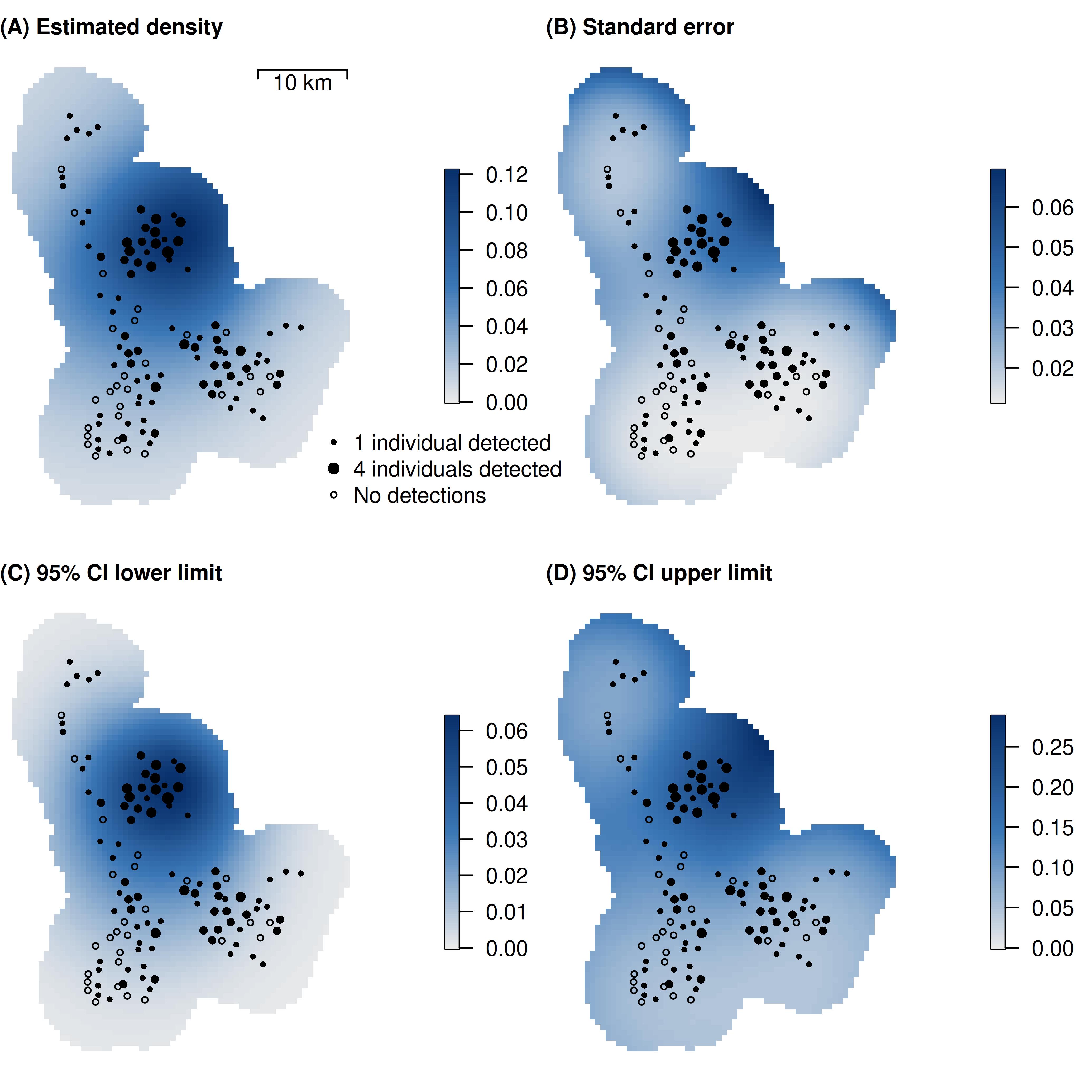}
	\caption{\textbf{Plot A:} Estimated density across the study area from the black bear data (individuals per km$^2$). Points correspond to hair-snare locations, with the size of the point indicating the number of different individuals detected there. \textbf{Plots B--D:} Standard errors, lower confidence interval limits, and upper confidence interval limits, respectively.}
	\label{fig:bear_results}
\end{figure}

The data used in this section are from a hair-snare survey of female black bears in Louisiana, USA, in 2007 \citep{chandler_SpatiallyExplicitIntegrated_2014}. The survey consisted of 115 hair-snares at which 39 unique individuals were detected over 8 survey occasions. Hair-snares are an example of a binary proximity detector: for each individual and on each occasion they record a binary indicator to indicate detection. Figure \ref{fig:bear_results} shows the detector locations, the number of individuals detected, and the total number of detections at each hair-snare. This data shows a clear spatial structure, with a higher number of individuals detected and total number of detections in the north as compared to the south and west of the study region.

We require a minor extension to the SCR model presented in Section \ref{sec:scr} in order to handle multi-occasion rather than single-occasion data. The slight modificatons are to $p_\cdot(\bm{s})$ and $f_{\bm{\omega} \mid \bm{s}}(\bm{\omega}_i \mid \bm{s}_i)$, which must include products over occasions \citep[see][for further details]{borchers_SpatiallyExplicitMaximum_2008}, but once these changes have been made the novelties we introduced to SCR in Section \ref{sec:scr-with-splines} remain the same. We fitted a density model with an intercept and a smooth effect of space to account for the clear spatial structure in the SCR data. For simplicity, no covariates were used to model detection parameters.

See Figure \ref{fig:bear_results}(A) for the estimated density over the study region. To obtain interval estimates we used a parametric bootstrap. We used the Hessian at the maximum likelihood estimate as a joint precision matrix under a Gaussian approximation of the likelihood to generate $10\,000$ realizations of the density surface. We extracted the standard deviation, along with the 0.025 and 0.975 quantiles, to obtain a standard error and 95\% confidence interval for each point in space.  The spatially varying density surface broadly reflects the spatial structure in the observed data, with higher estimated density in regions near traps that detected a larger number of individuals.

\cite{chandler_SpatiallyExplicitIntegrated_2014} used a homogeneous Poisosn point process within their SCR analysis of these data. They reported an estimate for the total number of female bears within the survey region in 2007 of $\widehat{N} = 50$, rounded to the nearest integer. The corresponding estimate under our model was also $\Lambda(\widehat{\bm{u}}) = 50$, where $\widehat{\bm{u}}$ are the spline coefficient values estimated by our model. This similarity reinforces previous findings \citep{efford_EstimatingPopulationSize_2013}: estimates of population size are robust to misspecifications of the density model within SCR. Although we fitted a more appropriate spatial model than \cite{chandler_SpatiallyExplicitIntegrated_2014} to account for inhomogeneous density, inference about total abundance is similar between the two. Our model, however, additionally provides appropriate inference about animal distribution, which is not available from those that fail to properly account for spatial features of the data.

\subsection{Application to gibbon acoustic survey data}
\label{sec:gibbons}

The southern yellow-cheeked crested gibbon \emph{Nomascus gabriellae} is a primate species listed as endangered by the IUCN \citep{rawson_GibbonRedList_2020}, and is threatened by habitat loss due to illegal logging, illegal poaching for food, and capture for the pet trade. In 2019, an acoustic survey was conducted in the Phnom Prich Wildlife Sanctuary (PPWS), a $2\,225$-km$^2$ protected area in Cambodia, to estimate the species' population density and distribution. Similar acoustic SCR surveys have been used for other primate populations \citep{Kidney_Gibbons_2015, mcgrath_Gibbons_2023, hankinson_Lar_2023}.

\begin{figure}
	\includegraphics[width = \textwidth]{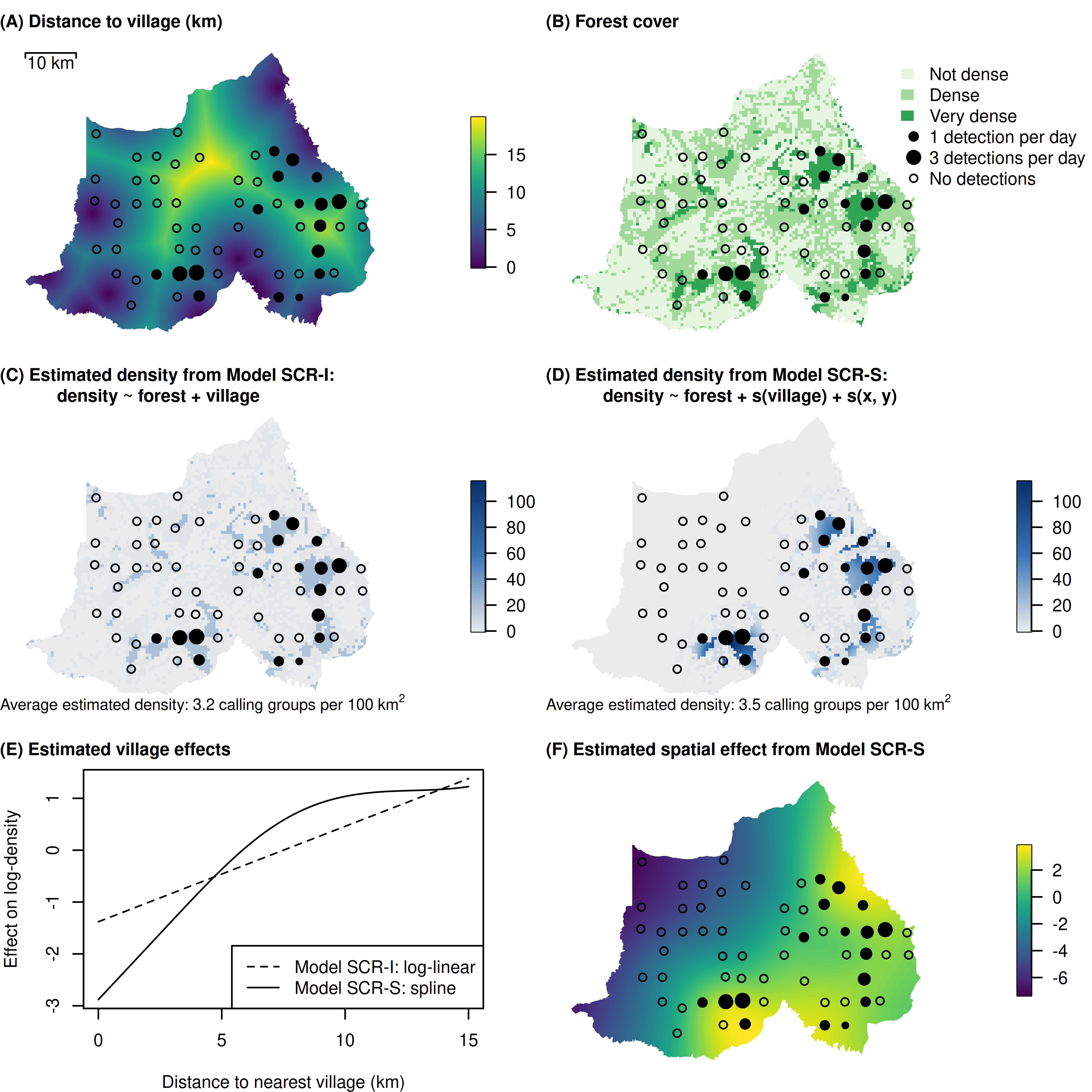}
	\caption{\textbf{Plots A--B:} Spatial covariates and locations of sets of listening posts. Point size indicates the average number of groups detected. \textbf{Plots C--D:} Estimated density (calling groups per 100 km$^2$) from Models SCR-I and SCR-S, respectively. \textbf{Plot E:} The estimated effects of distance to the nearest village under Models SCR-I and SCR-S. \textbf{Plot F:} The estimated spatial effect from Model SCR-S}
	\label{fig:gibbon_results}
\end{figure}

Each morning, detections of gibbon calls were recorded by surveyors stationed at three listening posts, which fell roughly in a straight line and separted by 500 m. Surveyors were stationed at the same listening posts for three consecutive days before moving to a new location. In total, data were collected at 58 separate sets of listening posts, providing good spatial coverage of PPWS (Figure \ref{fig:gibbon_results}). Spatial variables were collected to explain variation in gibbon density across PPWS, including forest type (very dense, dense, and not dense) and distance to the nearest village. Previous studies have used similar variables, including those related to forest cover and distance to human activity, to model spatial distribution of primates using acoustic SCR models \citep{mcgrath_Gibbons_2023, hankinson_Lar_2023}. These studies have demonstrated spatial variation amongst vegetation types, and that these primates avoid human activity.

Acoustic recorders (e.g., microphones, or in this case, human surveyors) are another example of binary proximity detector. Mathematically, acoustic SCR models operate in the same way as those described in Section \ref{sec:scr}, although the definitions for some of the variables and parameters involved are slightly different. Instead of $\bm{s}_i$ being the activity centre for the $i$th individual, it is the physical location of the $i$th vocalizing gibbon group. The detection function models the detectability of an acoustic signal as a function of physical distance between the gibbon group and the surveyor. The intensity function $D(\bm{s})$ provides the density of calling gibbon groups at location $\bm{s}$, rather than the density of all groups, because groups do not call every morning.

A minor extension to the SCR model presented in Section \ref{sec:scr} is also required here. On acoustic surveys, it is common for surveys to collect additional data that is informative about animal location. Here, each surveyor recorded an estimated bearing for each detection, so that $\bm{\omega}_i$ contains not only the binary detection responses, but also these estimated bearings. The estimated bearings are subject to error, however, and are assumed to arise from a von-Mises distribution centred on the true bearing. The concentration parameter of the von-Mises distribution is estimated by the model, and controls the magnitude of measurement errors. Under this model, $f_{\bm{\omega} \mid \bm{s}}(\bm{\omega}_i \mid \bm{s}_i)$ includes products of a Bernoulli PMFs and von-Mises PDFs. See \cite{Borchers_UnifyingModel_2015} for further details.

Here, we initially consider two SCR models that have previously been used for acoustic surveys of primates. The first we call SCR-H, with a homogeneous Poisson process for locations, similar to \cite{Kidney_Gibbons_2015}. The second we call SCR-I, with an inhomogeneous Poisson process for locations, using forest cover as a categorical covariate and a log-linear effect of distance to village as a numeric covariate, similar to \cite{mcgrath_Gibbons_2023}. We fit a third model, called SCR-S, using the methods described in this paper. In addition to using forest cover as a categorical covariate, this model includes a penalized regression spline to fit a smooth effect of distance to the nearest village, and another penalized regression spline to fit a smooth spatial effect. This model has two primary strengths relative to SCR-I. First, the log-linear relationship in SCR-I imposes a global effect of distance to the nearest village: moving 1 km father from a village produces the same change in log-density regardless of whether the location is close to or far from a village. We argue that a local effect is more realistic, because gibbons are likely to only respond to human disturbance near their territory: moving 1 km farther from a village may substantially increase density in areas close to villages, but may have little effect in areas that are already far away.  This is an example of the kind of biological realism that can be incorporated using our approach. The penalized regression spline in Model SCR-S can flexibly accommodate such a relationship. Second, Model SCR-S includes an LGCP fitted via a smooth spatial effect, thus accounting for variation in gibbon group density due to covariates that could not be observed (e.g., illegal logging activity), or due to clustering of gibbon groups that cannot be explained by the spatial covariates used in Model SCR-I.

The number of gibbon groups that were detected per day varied considerably across PPWS. The majority of detections were confined to the south and eastern regions of the park, and no listening posts in the northwest detected any gibbons (Figure \ref{fig:gibbon_results}). The homogeneous density model SCR-H provided a density estimate (with 95\% CI) of 6.3 (4.8, 7.9) calling gibbon groups per 100 km$^2$, but this model is clearly misspecified due to the spatial variation in the number of gibbon group detections. Under both SCR-I and SCR-S, calling gibbon group density was estimated to be substantially higher in very dense forest than the other forest types. However, inference about the effect of villages on calling gibbon group density differs between the two models. Under Model SCR-I, with a log-linear relationship, log-density is forced to increase linearly with increasing distance from village. However, inference from Model SCR-S suggests that the presence of a village only has a local effect out to 10 km, beyond which there is no effect of increasing distance (Figure \ref{fig:gibbon_results}E).

Moreover, Models SCR-I and SCR-S produced very different estimates of calling gibbon group distribution. Model SCR-I, without smooth effects, provides an estimated density of between 18--23 calling groups per 100 km$^2$ across the patches of very dense forest (depending on distance from village; Figure \ref{fig:gibbon_results}C), despite listening posts in some of these patches (e.g., the entire northwest region) detecting no gibbons. Model SCR-S included a smooth effect (Figure \ref{fig:gibbon_results}F), allowing flexibility for estimated density to vary across very dense forest. Consequently, there was a great deal more spatial variation in estimated density under Model SCR-S, providing a much better alignment with the observed data than SCR-I. Approximately 20\% of very dense forest was estimated to be virtually uninhabited (estimated density less than 1 calling group per 100 km$^2$), with high density of gibbons confined to pockets of very dense forest in the south and east (Figure \ref{fig:gibbon_results}D). The estimated distributions under Models SCR-I and SCR-S are so different that they would likely come with drastically different conservation implications and management decisions. Note that we can create standard errors and confidence intervals for these estimated density surfaces using the same bootstrap method from Section \ref{sec:bears}, but we omit them here for brevity.

Although the estimates of distribution are very different, estimates of average group density across PPWS are very similar: Models SCR-I and SCR-S provide average density estimates of 3.2 and 3.5 calling gibbon groups per 100 km$^2$, respectively, across PPWS. Estimated average density in very dense forest was also relatively similar for both models: 21 and 28 calling groups per 100 km$^2$, respectively. These results further reinforce previous findings that SCR models with oversimplified spatial effects nevertheless provide robust estimates of overall abundance \citep{efford_EstimatingPopulationSize_2013}.

\section{Simulation Study}
\label{sec:sim}

\begin{figure}[tbh]
	\includegraphics[width = \textwidth]{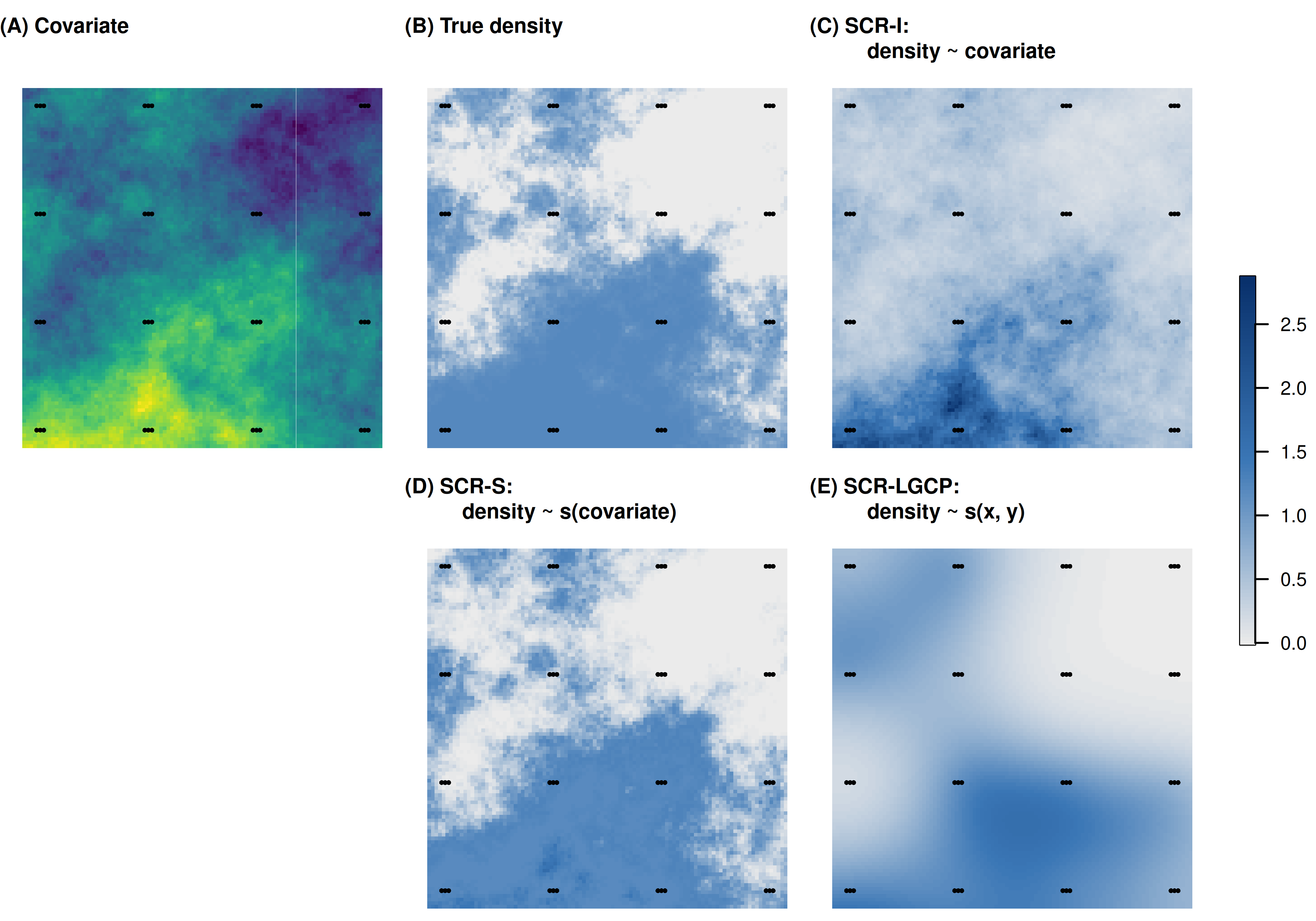}
	\caption{\textbf{Plot A:} The covariate. Points correspond to detector locations. \textbf{Plot B:} The true density surface that arises from Equation \eqref{eq:saturating-response}. \textbf{Plots C--E:} Per-pixel mean estimated density across all simulated data sets for Models SCR-I, SCR-S, and SCR-LGCP.}
	\label{fig:sim_D}
\end{figure}

We ran a simulation study to assess the performance of our proposed model against alternatives under known conditions. We simulated a spatial covarariate within a survery region, and constructed a density surface via a nonlinear function of this covariate. The relationship we selected is a saturating functional response, also known as a Type III Holling response \citep{holling_response_1959}. Increasing the covariate initially leads to an increase in animal density before levelling off at an asymptote, corresponding to the carrying capacity of the habitat. Let $z(\bm{s})$ denote the covariate at location $\bm{s}$. The density surface is $\log\{D(\bm{s})\} = \beta_0 + f\{z(\bm{s}\}$, where the effect of the covariate is given by
\begin{equation}
	f(z) = \alpha [1 - \exp\{-\gamma(z - \tau)\}]. \label{eq:saturating-response}
\end{equation}
We set parameters $\alpha = 1$, $\gamma = 3$ and $\tau = -2$. In Figure \ref{fig:sim_D}, we plot the covariate, the resulting intensity function, and the true $f(z)$ as a purple line.

We compared the performance of four SCR models: (1) SCR-H, with a homogeneous Poisson process; (2) SCR-I, with an inhomogeneous Poisson process modelled via a log-linear effect of the covariate; (3) SCR-S, with a smooth effect for the covariate modelled via a one-dimensional penalized regression spline; and (4) SCR-LGCP, with a smooth effect over space modelled via a two-dimensional penalized regression spline. Models SCR-H and SCR-I are already widely used SCR models, but are misspecifications here. Models SCR-S and SCR-LGCP do not explicitly include the parametric relationship in Equation \eqref{eq:saturating-response}, but they include penalized splines to flexibly estimate $f(z)$ (Model SCR-S), or the resulting animal density surface $D(\bm{s})$ without using covariates (Model SCR-LGCP). We used thin plate regression splines for both Models SCR-S and SCR-LGCP.

Our survey scenario was inspired by the acoustic survey in Section \ref{sec:gibbons}. We arranged the detectors in a regular pattern across the spatial domain in clusters of three. For simplicity, we used a single survey occasion with binary detections, where $\omega_{ij} = 1$ if individual $i$ is detected on trap $j$, and zero otherwise. In addition to the binary detection data, we also simulated estimated bearings for detected calls from a von-Mises distribution with a similar concentration parameter as that estimated from the gibbon data. The detection function was also selected to roughly match that estimated from the gibbon data: a half-normal function with $g_0 = 1$ and $\sigma = 1500$.

\begin{figure}[tbh]
	\includegraphics[width = \textwidth]{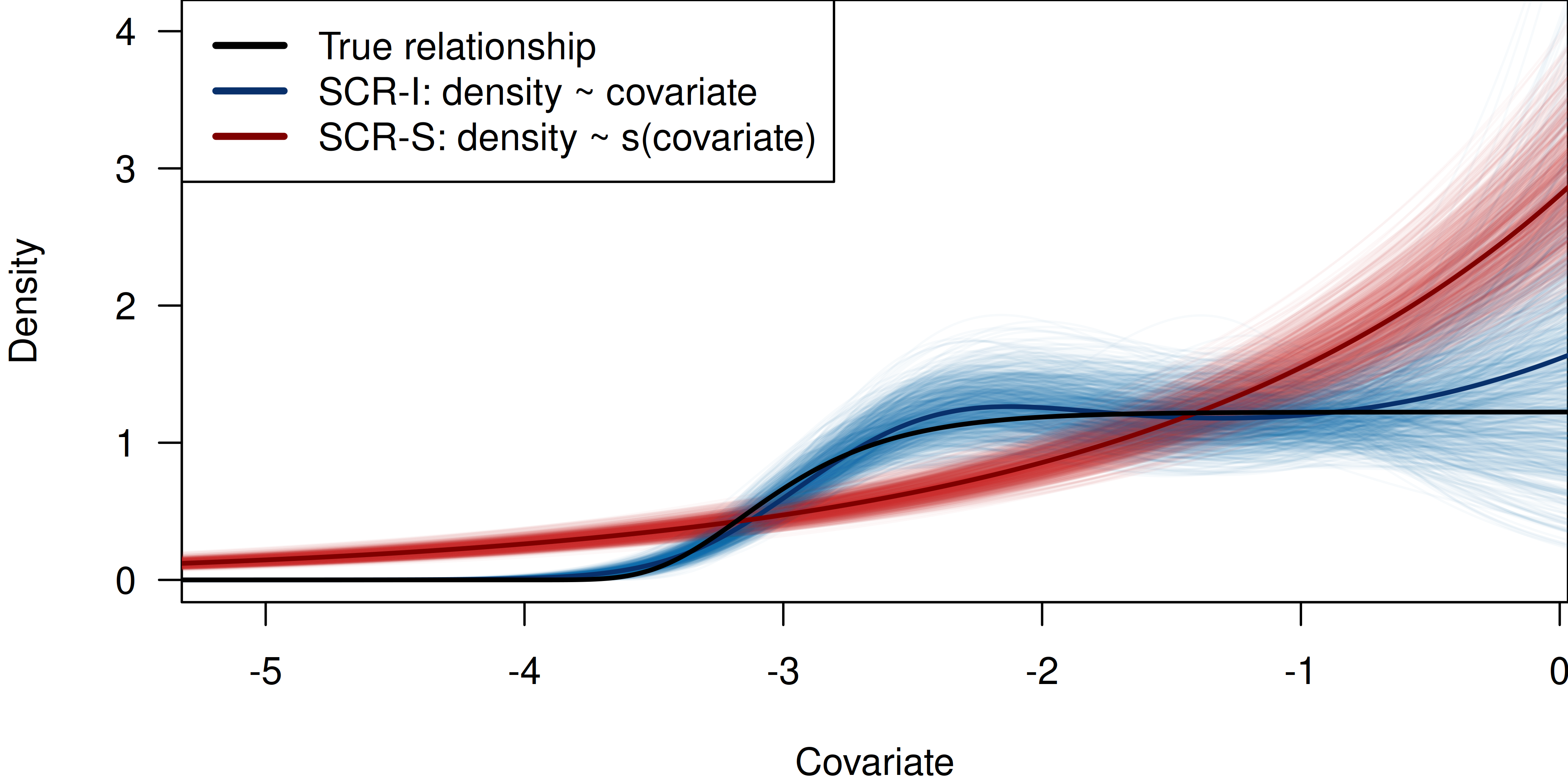}
	\caption{Estimated relationships between the covariate and density from Models SCR-I and SCR-S. Each individual transparent line is a relationship estimated from a single data set, while the darker solid line is the average across all simulated data sets.}
	\label{fig:sim_cov}
\end{figure}

In Figure \ref{fig:sim_cov}, we plot the estimated relationships between the covariate and animal density from Models SCR-I and SCR-S for each simulated data set. On average, the nonlinear relationship estimated by Model SCR-S using a penalized spline is close to the true function $f(z)$. The log-linear relationship in Model SCR-I is incorrect a misspecification, and consequently the model is forced to balance overprediction (at low and high values of the covariate) and underprediction (at middling values).

Figure \ref{fig:sim_D} shows the per-pixel average estimated intensity across all simulations. Model SCR-S, with the penalized regression spline applied to the covariate, most closely matches the true intensity surface. Although it does not explicitly include the parametric relationship in Equation \eqref{eq:saturating-response}, the penalized spline can flexibly capture this nonlinear function. As per Figure \ref{fig:sim_cov}, Model SCR-I overpredicts at locations with low or high covariate values, and underpredicts at locations with middling covariate values.

Model SCR-LGCP, with the smooth over space, has captured the large-scale trends in the intensity surface. Without covariate information it is difficult to infer fine-scale variation in activity centre locations, especially for SCR where the locations are latent and we only observe detection data. The overprediction in the upper left corner of the region is because the three clusters of detectors in that region (in the top-left corner, and those vertically and horizontally adjacent) have, by chance, been placed in local pockets of higher relative density. Since the model has no information about the regions between clusters of detectors, it has interpolated between these higher density regions to infer a general higher density in the vicinity, which is common behaviour for interpolation with spatial effects. To address this in practice we would need a higher density of detector clusters. Nevertheless, the model has performed well in terms of estimating large-scale spatial variation in density across most of the region.

\begin{figure}[tbh]
	\includegraphics[width = \textwidth]{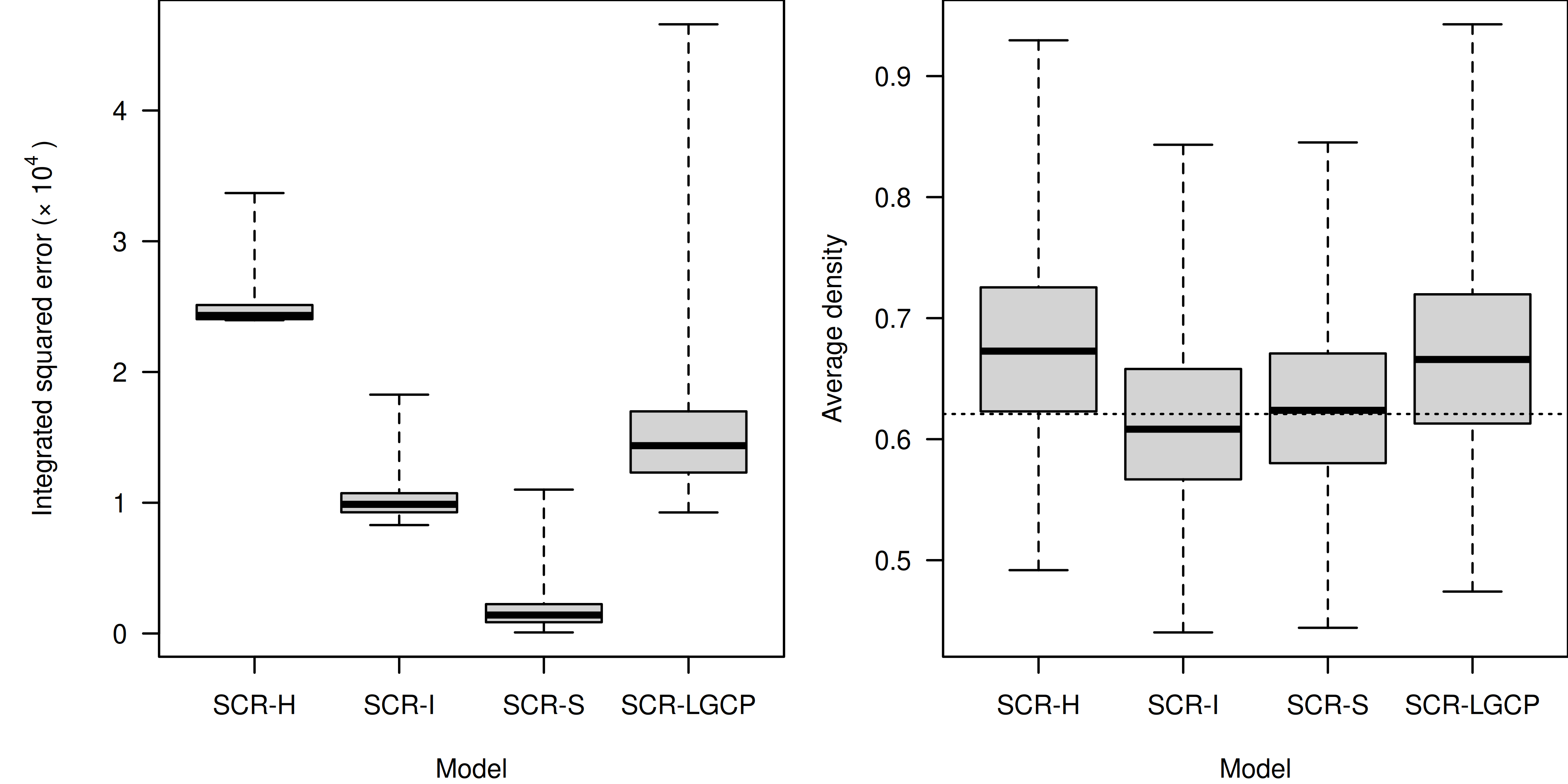}
	\caption{A summary of density estimate performance for all models across the simulated data sets, with integrated squared error (left) and average estimated density (right).}
	\label{fig:sim-boxplots}
\end{figure}

To assess the difference between the estimated intensity surface and the truth, we calculated the integrated squared error (ISE) for each model.  Figure \ref{fig:sim-boxplots} (left) summarises these ISE values with a boxplot for each model. As expected, Model SCR-S has the lowest ISE, because it includes covariate information (unlike Models SCR-H and SCR-LGCP) and can flexibly approximate the true function $f(z)$ (unlike Model SCR-I).

The second-best model in terms of ISE is Model SCR-I, the inhomogeneous model with a log-linear effect of the covariate. We found this result slightly surprising because it overestimates density quite severely in certain regions due to the misspecified effect of the covariate (Figure \ref{fig:sim_D}. However, unlike Model SCR-LGCP, this model does have access to the spatial covariate $z(\bm{s})$. Model SCR-LGCP, with a two-dimensional penalized spline over space, does not have access to the fine-scale information held in the covariate, and performs slightly worse by ISE. However, it is a noticeable improvement over Model SCR-H, and is close in performance to Model SCR-I, despite ignoring the covariate information.

We also checked the ability of each model to estimate the average density across the whole region, which is often a target of inference for SCR. We observe positive bias from the models that do not use the covariate, Models SCR-H ($9.3\%$ bias) and SCR-LGCP ($7.8\%$), because, on average, density surrounding the $16$ cluters of detectors was slightly higher than density throughout the entire region. Model SCR-S is roughly unbiased ($1.2\%$ bias), while SCR-I, with the misspecified effect of the covariate, has slight negative bias ($-1.1\%$).

\section{Discussion}

We have developed a new SCR method that allows wildlife population density to be modelled using penalized regression splines. Our applications and our simulation study demonstrate that our approach captures broad-scale variation in animal density without relying on environmental covariates by fitting smooth functions over space. Meanwhile, our application to gibbon data, alongside our simulation study, similarly demonstrate that we can flexibly model nonlinear relationships between spatial covariates: in Section \ref{sec:gibbons} the estimated effect of distance from village flattens off at approximately 10 km (Figure \ref{fig:gibbon_results}), consistent with local effects of villages on gibbon density, while in Section \ref{sec:sim} the average estimated relationship closely follows the true Type III Holling response (Figure \ref{fig:sim_cov}).

We focused on Laplace-approximate maximum likelihood for inference, although we expect MCMC is one feasible alternative. We note that development of Bayesian approaches to estimating inhomogeneous density surfaces in SCR has been slow: although maximum likelihood methods were initially proposed by \cite{borchers_SpatiallyExplicitMaximum_2008}, fast and efficient MCMC samplers to fit SCR models with inhomogeneous Poisson processes have only been developed relatively recently \citep{zhang_ihp_2023}. Bayesian approaches to fitting more sophisticated spatial processes within SCR is therefore a potential avenue for future research.

We highlight that care must be taken when interpreting estimated intensity surfaces from LGCPs, regardless of whether they are embedded within an SCR model (like we consider here), or if the LGCP is fitted directly to point pattern data. When inference is based on a single realisation of the point process, it is impossible to separate spatial variation in the estimated intensity due to systematic effects that persist across multiple realizations (epistemic uncertainty) and spatial variation due to processes that are inherently stochastic (aleatoric uncertainty). By way of example, in our black bear case study we identified higher density of black bears near the north of the study area. This hot-spot in density may be due to good-quality habitat in this region (e.g., abundant resources), in which case we would expect a hot-spot in the same location under new hypothetical realizations of activity centre locations. Alternatively, this hot-spot may have arisen because animals tend to cluster together for social reasons, in which case hypothetical realizations may have similar-sized hot-spots, but they may arise in different locations. Interpreting spatial variation in estimated intensity surfaces from LGCPs therefore relies on additional information, such as domain knowledge or data collected over longer time periods rather than a single snapshot in time.

This difficulty in interpretation can cause some to make claims such as the view that it is impossible to fit a LGCP to a single realisation, and that all that is achieved in the attempt is fitting an IHPP which includes some penalised splines.  A riposte would be that an IHPP with random effects is the definition of a Cox process.  These differences in opinion result from a philosophical difference of perspective and, perhaps, a lack of understanding of the relationship between splines and Gaussian random fields.  

For some, the only way to distinguish between LGCPs and IHPPs when working with a single realisation is for researchers to ask themselves what the model means to them.  What would I think \textit{were I to observe another realisation of the process}?  Under a single realisation of the process, this may be the only distinguishing feature that allows a researcher to decide if they have fitted an IHPP or a Cox process model, particularly if one is working in a Bayesian context where all parameters are random.  Others would reject allowing such subjectivism to so strongly define the nomenclature of a modelling paradigm. We would emphasise that this debate does not matter nearly as much as the mere ability to flexibly model the effect of covariates using penalised splines, regardless of the wider point process interpretation of such an act.

To conclude, our proposed method provides a flexible way to model spatial variation in wildlife population density, particularly in surveys that span large and heterogeneous regions. Many SCR surveys fall into this category: camera-trap and genetic surveys often span vast geographic landscapes \citep[e.g.,][]{bischof_transnational_monitoring_2020}, and the decreasing cost of acoustic recorders makes it feasible to deploy clusters of detectors across large regions, similar to the configuration of listening posts for the gibbon survey described in Section \ref{sec:gibbons}. By estimating density and distribution of wildlife populations in these settings, our approach offers a useful addition to the suite of tools available for conservation and wildlife management.

\section*{Acknowledgements}

We thank the Ministry of Environment, Royal Government of Cambodia for permitting and supporting the study in Phnom Prich Wildlife Sanctuary. We are grateful for the logistical and financial support to the field work from WWF Cambodia, and to the community researchers for their dedicated role in data acquisition.

We are grateful to R. B. Chandler, J. D. Clark, C. Lowe, K. O’Connell and the Louisiana Department of Wildlife and Fisheries for graciously allowing the use of the Louisiana black bear data.

\section*{Data Availability Statement}

The gibbons data and code to reproduce the results in this paper can be accessed at \url{https://anonymous.4open.science/r/scr-lgcp-public-CCA1/README.md}.  We are currently in discussions around sharing the black bear data and will update this section when these are resolved.  As a place holder we have also included a pseudo-dataset based on a simulation and the code will produce results very similar to our analysis of the real black bear data.  

\section*{Disclosure statement}

None of the authors have any known conflict of interest in relation to this work.

\bibliography{paper}

\end{document}